\documentclass{ws-procs975x65}

\begin{document}



\title{Impact of dark matter on reionization and heating}

\author{MICHELA MAPELLI}

\address{SISSA/ISAS,
Via Beirut 2-4, 
Trieste I-34014, Italy 
, \email{mapelli@sissa.it}}

\author{EMANUELE RIPAMONTI}

\address{Kapteyn Astronomical Institute, University of Groningen, Postbus 800, 9700 AV,\\ Groningen, The Netherlands, \email{ripa@astro.rug.nl}}


\begin{abstract}
We derived the evolution of the energy deposition in the intergalactic medium (IGM) by different decaying (or annihilating) dark matter (DM) candidates. Heavy annihilating DM particles (with mass larger than a few GeV) have no influence on reionization and heating, even if we assume that all the energy emitted by annihilations is absorbed by the IGM. In the case of lighter particles, the impact on reionization and heating depends on the efficiency of energy absorption by the IGM. We calculated the fraction of energy produced by decays and annihilations which is effectively absorbed by the IGM. We found that this fraction is generally high at very high redshift ($\gg{}100$), but drops at more recent epochs.
\end{abstract}

\bodymatter

\section{Introduction}\label{intro}

The reionization and heating history of the intergalactic medium (IGM) has not yet been fully understood. In particular, the nature of the sources of these processes is mostly unclear. An important contribution might arise from the first stars; but other more exotic sources (e.g. intermediate mass black holes, decaying dark matter particles, etc.) have been proposed as well.\cite{CF}
In this proceeding, we focus on the possible role of dark matter (DM) decays and annihilations on reionization and heating of the IGM.

\section{The energy absorbed fraction}
The rate of energy released by DM decays/annihilations and absorbed (per baryon) by the IGM can be expressed as
\begin{equation}\label{eq:fabs}
\epsilon{}(z)=f_{\rm abs}(z)\,{}\dot{n}_{\rm DM}(z)\,{}m_{\rm DM}c^2,
\end{equation}
where $f_{\rm abs}(z)$ is the fraction of energy absorbed by the IGM, $m_{\rm DM}$ is the mass of the DM particle and $\dot{n}_{\rm DM}(z)$ is the decrease rate of the number of DM particles per baryon. The expression of  $\dot{n}_{\rm DM}(z)$ depends on the density of DM particles and on the decays lifetime or on the thermally averaged annihilation cross-section, in the case of decays and annihilations, respectively.\cite{RMF06} 

The most crucial parameter in equation~(\ref{eq:fabs}) is $f_{\rm abs}(z)$. Previous studies\cite{MFP06} assumed a complete and immediate energy absorption by the IGM (i.e. $f_{\rm abs}(z)=1$), deriving upper limits to the effects of DM on reionization and heating. Annihilating heavy DM particles ($m_{\rm DM}c^2{}\gtrsim{}$ GeV) have no effects on such processes, even assuming $f_{\rm abs}(z)=1$, because of the very small allowed interacting rate. Instead, under the assumption of complete absorption, lighter particles ($m_{\rm DM}c^2{}\lesssim{}$100 MeV) could be important sources of partial early reionization and heating\cite{MFP06}.

$f_{\rm abs}(z)$ can be hardly calculated in the case of heavy particles ($m_{\rm DM}c^2{}\gg$100~MeV), because of the
uncertainties in modeling the cascade associated with massive product particles\footnote{Modeling the cascade is not only difficult, but also not necessary for our purposes, since the effects of massive particles on reionization are negligible\cite{MFP06}, even if we assume $f_{\rm abs}(z)=1$.}.
On the other hand, for relatively light ($m_{\rm DM}c^2\lesssim{}100$ MeV) DM candidates it is possible to derive the correct behaviour of the absorbed fraction\cite{RMF06} $f_{\rm abs}(z)$.
In fact, the possible decay/annihilation
products of these light particles are only photons, electron-positron pairs, or neutrinos (which are
assumed to have negligible interactions with matter).
For photons  the effects of Compton scattering and photo-ionization must be considered; for
pairs, the relevant processes are inverse Compton scattering,
collisional ionizations, and positron annihilations.

If the decay products are both photons and pairs, $f_{\rm abs}(z)$ is found to be high at early epochs ($z\gg{}100$), when  photo-ionization (considering photons) and  inverse Compton scattering or/and positron annihilations (considering pairs) are efficient processes. In both cases, $f_{\rm abs}(z)$ drops at lower redshifts. As an example, in Fig.~1 we show the behaviour of $f_{\rm abs}(z)$ in the case of sterile neutrinos which decay into photons (left panel) and light dark matter (LDM) particles which annihilate producing pairs (right panel). Similar results are obtained for decaying LDM particles\cite{RMF06}.

\def\figsubcap#1{\par\noindent\centering\footnotesize(#1)}

\begin{figure}\label{fig:fig1}%
\begin{center}
 \parbox{2.3in}{\epsfig{figure=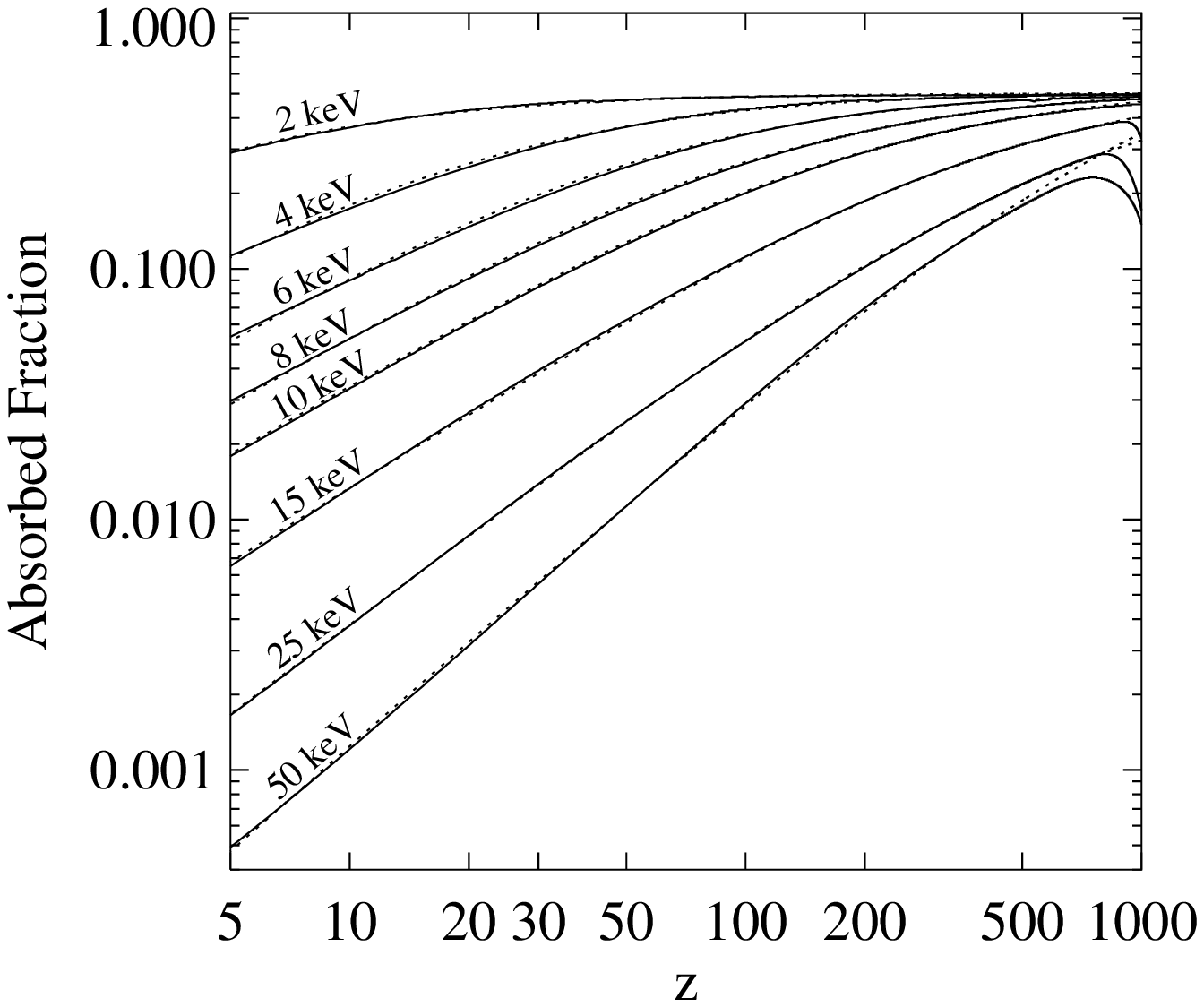,width=2.3in}
 \figsubcap{a}}
 \hspace*{4pt}
 \parbox{2.3in}{\epsfig{figure=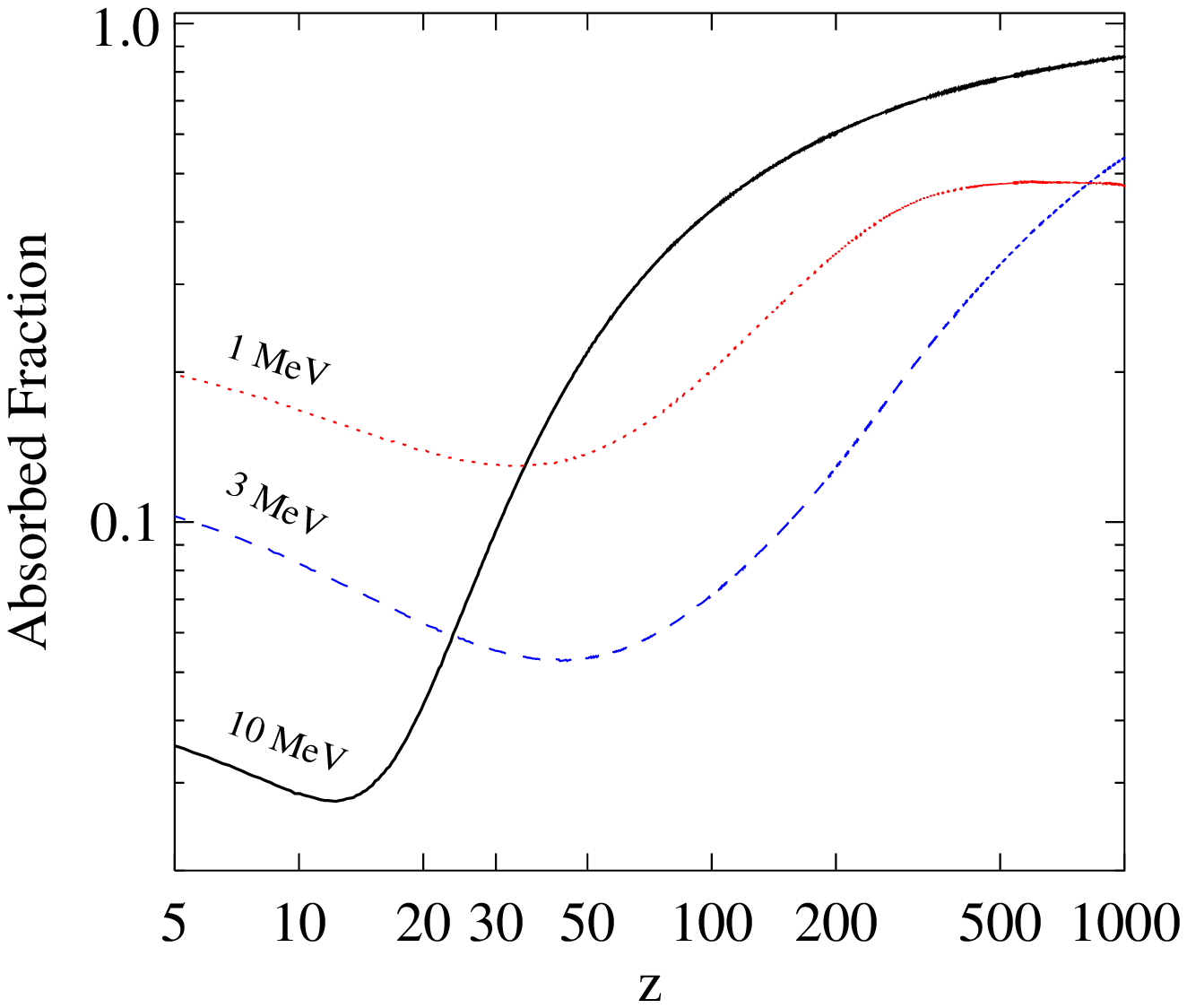,width=2.3in}
 \figsubcap{b}}
 \caption{Absorbed fraction as a function of redshifts for sterile neutrino decays (a) and LDM annihilations (b).}
\end{center}
\end{figure}



\section{Effects of decays and annihilations on reionization and heating}
We can now calculate how the behaviour of $f_{\rm abs}(z)$ affects the impact of DM decays/annihilations on reionization and heating\cite{RMF06}. As an example, in Fig.~2 we again consider the case of sterile neutrinos decaying into photons (left panel) and of LDM particles annihilating into pairs (right panel). The thin and thick lines in both panels refer to the case where we assume complete absorption and to the case in which  $f_{\rm abs}(z)$ is the same as shown in Fig.~1, respectively. The effect of sterile neutrino decays  on reionization and heating is strongly suppressed, if we consider the correct $f_{\rm abs}(z)$. A similar result can be found for LDM decays\cite{RMF06}. Also in the case of LDM annihilations this suppression is apparent, even if less dramatic.
\begin{figure}[b]\label{fig:fig2}%
\begin{center}
 \parbox{2.325in}{\epsfig{figure=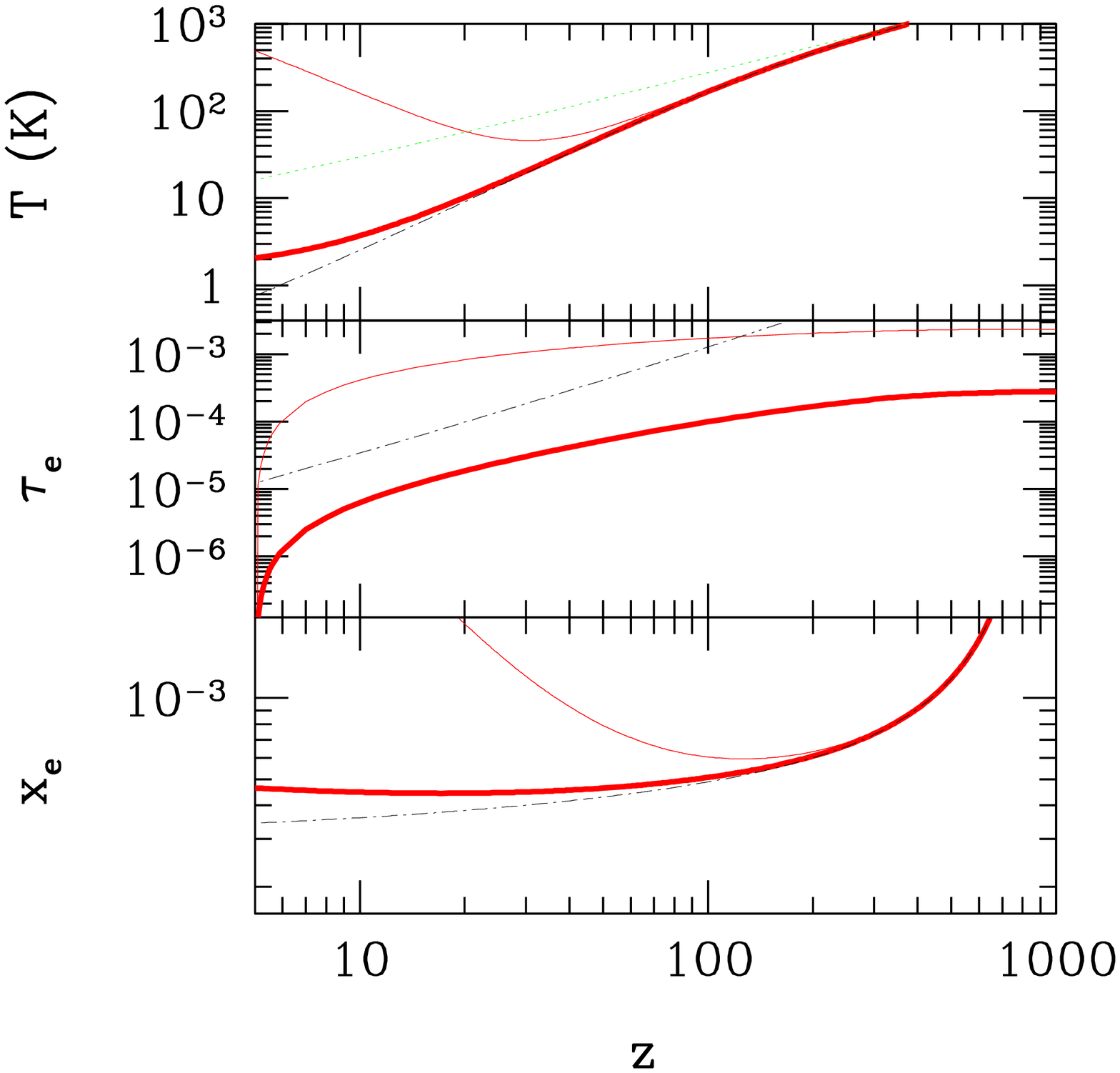,width=2.325in}
 \figsubcap{a}}
 \hspace*{4pt}
 \parbox{2.325in}{\epsfig{figure=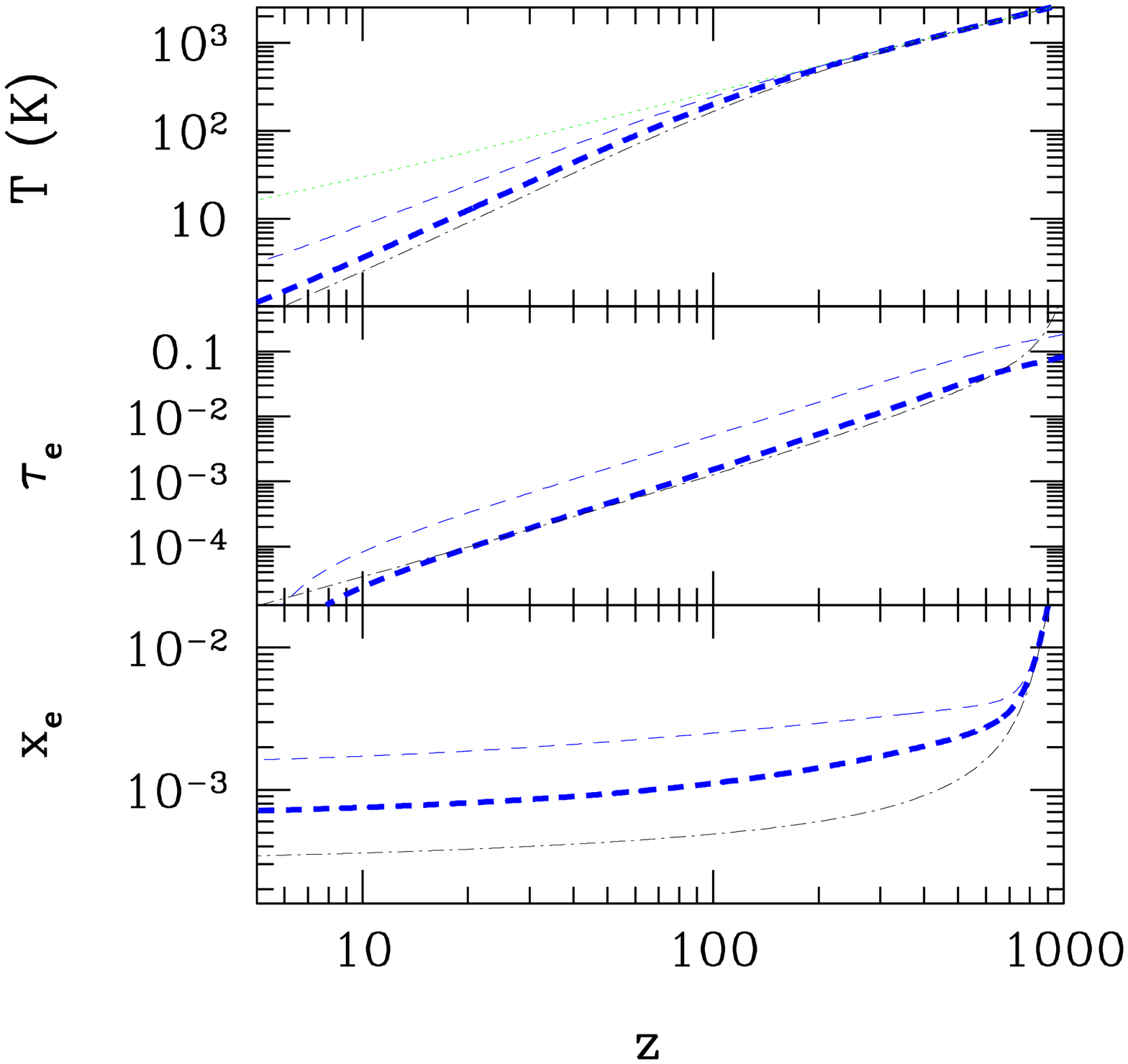,width=2.325in}
 \figsubcap{b}}
 \caption{Ionized fraction (bottom panels), Thomson optical depth (central panels) and IGM temperature (top panels) as a function of redshift due to sterile neutrino decays (a) and LDM annihilations (b). 
The thin dot-dashed
line represents, from bottom to top, the relic fraction of free
electrons, their contribution to Thomson optical depth and the IGM
temperature without particle decays.  In the top panel, the thin dotted
line represents the CMB temperature.
In the left (right) panel the thin and the thick solid (dashed) lines refer to 25-keV sterile neutrino decays (3-MeV LDM annihilations), in the hypothesis of complete energy absorption and taking into account the effective absorbed fraction $f_{\rm abs}(z)$ (Fig. 1), respectively.
}
\end{center}
\end{figure}

In summary, we can conclude that the correct calculation of $f_{\rm abs}$ is crucial\cite{RMF06} : if the correct values of $f_{\rm abs}(z)$ are taken into account,  the impact of DM decays and annihilations on reionization and heating is almost negligible, a factor $\sim{}2$ to 1000 lower than previous estimates based on the hypothesis of complete and immediate absorption.

\section*{Acknowledgments}
MM acknowledges the organizers of the Eleventh Marcel Grossmann meeting for the MGF grant. The authors thank P. L. Biermann for inviting them to the meeting. ER acknowledges support from NWO grant 436016.


\vfill

\end{document}